%% file: main.tex
\documentclass[letterpaper,twocolumn]{article}
\usepackage[margin=.8in]{geometry}
\usepackage{natbib}
\usepackage{graphicx}
\usepackage{amsmath}
\usepackage{amssymb}
\usepackage{listings}
\usepackage{xcolor}
\DeclareMathOperator{\expect}{\mathbb{E}}
\DeclareMathOperator{\diag}{diag}
\DeclareMathOperator{\argmax}{argmax}

\setkeys{Gin}{width=.8\linewidth}
\graphicspath{{./images/}}
\usepackage{authblk}
\usepackage{times}
\usepackage{url}

\title{Joint-perturbation simultaneous pseudo-gradient}

\author[1]{Carlos Martin}
\author[1,2,3,4]{Tuomas Sandholm}
\affil[ ]{\{cgmartin, sandholm\}@cs.cmu.edu}
\affil[1]{Carnegie Mellon University}
\affil[2]{Strategy Robot, Inc.}
\affil[3]{Optimized Markets, Inc.}
\affil[4]{Strategic Machine, Inc.}
\date{}

\begin{document}
\maketitle
\input{abstract}
\input{introduction}
\input{related_research}
\input{formulation}
\input{method}
\input{experiments/main}

\input{conclusion}
\appendix
\input{acknowledgements}
\bibliographystyle{plainnat}
\bibliography{dairefs,references}
\input{code}
\end{document}

%% file: abstract.tex
\begin{abstract}
We study the problem of computing an approximate Nash equilibrium of a game whose strategy space is continuous without access to gradients of the utility function.
Such games arise, for example, when players' strategies are represented by the parameters of a neural network.
Lack of access to gradients is common in reinforcement learning settings, where the environment is treated as a black box, as well as equilibrium finding in mechanisms such as auctions, where the mechanism's payoffs are discontinuous in the players' actions.
To tackle this problem, we turn to zeroth-order optimization techniques that combine pseudo-gradients with equilibrium-finding dynamics.
Specifically, we introduce a new technique that requires a number of utility function evaluations per iteration that is constant rather than linear in the number of players.
It achieves this by performing a single joint perturbation on all players' strategies, rather than perturbing each one individually.
This yields a dramatic improvement for many-player games, especially when the utility function is expensive to compute in terms of wall time, memory, money, or other resources.
We evaluate our approach on various games, including auctions, which have important real-world applications.
Our approach yields a significant reduction in the run time required to reach an approximate Nash equilibrium.
\end{abstract}

%% file: introduction.tex
\section{Introduction}
\label{sec:introduction}

We tackle the problem of computing an approximate Nash equilibrium of a game with a black-box utility function, for which we lack access to gradients.
A standard way to learn player strategies for a game is to use simultaneous gradient ascent, in which, at each iteration, each player myopically adjusts its parameters to increase its own utility, treating the other players as fixed.
Computing the simultaneous gradient requires taking gradients of the utility function, which are unavailable in the black-box setting.
To address this obstruction, one can employ evolution strategies, a family of methods that perturbs parameters and evaluates the function at those perturbed points in order to optimize some objective.
This yields an unbiased estimator of the gradient of a smoothed version of the original objective function, also called a \emph{pseudo-gradient}.
Computing the simultaneous pseudo-gradient through the standard approach requires a number of utility function evaluations that is \emph{linear} in the number of players.
Our contribution is to introduce a new method which requires a number of function evaluations that is only \emph{constant} in the number of players.
It performs a joint perturbation on \emph{all} players' strategies at once, rather than perturbing each one individually.
When utility function evaluation is expensive (in terms of wall time, memory, money, \emph{etc.}) and the number of players is large, this can yield dramatic benefits for training.
We benchmark our approach on several games, including various auctions, showing a significant reduction in training time.

In \S\ref{sec:related_research}, we describe related research.
In \S\ref{sec:formulation}, we introduce relevant notation and present a mathematical formulation of the problem we are solving.
In \S\ref{sec:method}, we present our method.
In \S\ref{sec:experiments}, we present our experimental settings, results, and discussion.
In \S\ref{sec:conclusion}, we present our conclusions.

%% file: related_research.tex
\section{Related research}
\label{sec:related_research}

Black-box zeroth-order optimization uses only function evaluations to optimize a black-box function with respect to a set of inputs.
In particular, it does not require gradients.
There is a class of black-box optimization algorithms called \emph{evolution strategies (ES)} \citep{Rechenberg_1973, Schwefel_1977, Rechenberg_1978, back_1996, back_1997, eiben_2003}.
These maintain and evolve a population of parameter vectors.
\emph{Natural evolution strategies (NES)} \citep{Wierstra_2008, Sun_2009, Wierstra_2014} represent the population as a \emph{distribution over parameters} and maximize its average objective value using the score function estimator.
For many parameter distributions, such as Gaussian smoothing, this is equivalent to evaluating the function at randomly-sampled points and estimating the gradient as a sum of estimates of directional derivatives along random directions \citep{fu2015handbook, Duchi_2015, Nesterov_2017, Shamir_2017, Berahas_2022}.

\citet{li2021evolution} tackle the problem of solving symmetric one-shot Bayesian games with no given analytic structure, high-dimensional type and action spaces, many players, and general-sum payoffs.
They represent agent strategies in parametric form as neural networks, and apply NES to optimize them.
For pure equilibrium computation, they formulate the problem as bi-level optimization and use NES to implement both inner-loop best response optimization and outer-loop regret minimization.
For mixed equilibrium computation, they adopt an incremental strategy generation framework in which NES produces a finite sequence of approximate best-response strategies.
They then calculate equilibria over this finite strategy set via a model-based optimization process.
Both methods use NES to search for strategies over the functional space of policies, given only black-box simulation access to noisy payoff samples.

To tackle symmetric auctions, \citet{bichler2021learning} present a learning method called \emph{neural pseudogradient ascent (NPGA)} that represents strategies as neural networks and applies policy iteration on the basis of gradient dynamics in self-play to provably learn local equilibria.
The method follows the simultaneous gradient of the game and uses a smoothing technique to circumvent discontinuities in the \emph{ex post} utility functions of auction games.
(In auctions, discontinuities arise at the bid value where an arbitrarily small change makes the difference between winning and not winning.)
The method converges to a Bayesian Nash equilibrium for a wide variety of symmetric auction games.

Whereas symmetric auction models are widespread in the theoretical literature, in most auction markets in the field, one can observe different classes of bidders having different valuation distributions and strategies.
Asymmetry of this sort is almost always an issue in real-world multiobject auctions, in which different bidders are interested in different packages of items.
Such environments require a different implementation of NPGA with multiple interacting neural networks having multiple outputs for the different allocations in which the bidders are interested.
\citet{bichler2023learning} analyze a wide variety of asymmetric auction models.
Their results show that they closely approximate Bayesian Nash equilibria in all models in which the analytical Bayes–Nash equilibrium is known.
Additionally, they analyze new and larger environments for which no analytical solution is known and verify that the solution found approximates equilibrium closely.

\citet{bichler2023computing} introduce an algorithmic framework for Bayesian games with continuous type and action spaces, such as auctions.
It discretizes the type and action spaces and then learns distributional strategies \citep{Milgrom85:Distributional} (a form of mixed strategies for Bayesian games) via online convex optimization, specifically \emph{simultaneous online dual averaging (SODA)}.
They show that the equilibrium of the discretized game approximates an equilibrium in the continuous game.
Discretization can work well for small games, but does not scale to high-dimensional observation and action spaces.

\citet{ijcai2023p317} study the problem of computing an approximate Nash equilibrium of continuous-action game without access to gradients.
They model players' strategies as artificial neural networks.
In particular, they use randomized policy networks to model mixed strategies.
These take noise in addition to an observation as input and can flexibly represent arbitrary observation-dependent, continuous-action distributions.
Being able to model such mixed strategies is crucial for tackling continuous-action games that lack pure-strategy equilibria.
They apply this method to continuous Colonel Blotto games, single-item and multi-item auctions, and a visibility game, showing that it can quickly find a high-quality approximate equilibrium.

%% file: formulation.tex
\section{Formulation}
\label{sec:formulation}

Throughout the paper, we use the following notation.
The operator \(\otimes\) denotes the tensor product.
The operator \(\odot\) denotes the elementwise a.k.a. Hadamard product.
If \(\mathcal{S}\) is a set, \(\triangle \mathcal{S}\) is the set of all Borel probability measures on \(\mathcal{S}\).
If \(\mathcal{F}\) is a family indexed by \(\mathcal{I}\), \(\mathcal{F}_\times = \prod_{i \in I} \mathcal{F}_i\).
If \(\mu_i \in \triangle \mathcal{S}_i\) is a probability measure for \(i \in \mathcal{I}\), \(\bigotimes_{i \in \mathcal{I}} \mu_i \in \triangle \mathcal{S}_\times\) is the product measure.

\paragraph{Strategic-form game.}
A strategic-form game is a tuple \((\mathcal{I}, \mathcal{S}, u)\) where \(\mathcal{I}\) is a set of players, \(\mathcal{S}_i\) is a set of strategies for \(i \in \mathcal{I}\), and \(u : \mathcal{S}_\times \to \mathbb{R}^\mathcal{I}\) is a utility function.
A strategy profile is an element of \(\mathcal{S}_\times\), that is, an assignment of a strategy to each player.
The notation \(s_{-i}\) denotes \(s\) excluding \(i \in \mathcal{I}\).
Given a strategy profile, a \emph{best response (BR)} for a player is a strategy that maximizes its utility given the other players' strategies.
That is, given \(s \in \mathcal{S}_\times\), a BR for \(i \in \mathcal{I}\) is an element of \(\mathcal{B}_i(s_{-i}) = \argmax_{r_i \in \mathcal{S}_i} u(r_i, s_{-i})_i\).
A \emph{Nash equilibrium (NE)} is a strategy profile for which each player's strategy is a BR to the other players' strategies.
That is, it is an \(s \in \mathcal{S}_\times\) such that \(s_i \in \mathcal{B}_i(s_{-i})\) for all \(i \in \mathcal{I}\).

\paragraph{Exploitability.}
Given \(s \in \mathcal{S}_\times\), the BR utility of \(i \in \mathcal{I}\) is \(b_i(s_{-i}) = \sup_{r_i \in \mathcal{S}_i} u_i(r_i, s_{-i})\), \emph{i.e.}, the highest utility it could attain by unilaterally changing its strategy.
Player \(i\)'s regret is \(R_i(s) = b_i(s_{-i}) - u_i(s)\), which is the highest utility it could \emph{gain} from unilaterally changing its strategy.
The \emph{exploitability} is defined as \(\Phi = \sum_{i \in \mathcal{I}} R_i\).
It is non-negative everywhere and zero precisely at NE.
Consequently, it is used as a standard measure of ``closeness'' to NE in the literature \citep{Lanctot17:Unified, Lockhart_2019, Walton_2021, Timbers_2022}.
If NE exist, computing them is equivalent to globally minimizing exploitability.

\paragraph{Mixed strategies.}
For any game \((\mathcal{I}, \mathcal{S}, u)\), there exists a mixed-strategy game \((\mathcal{I}, \Sigma, \bar{u})\) where \(\Sigma_i = \triangle \mathcal{S}_i\) and \(\bar{u}_i(\sigma) = \expect_{s \sim \bigotimes_{j \in \mathcal{I}} \sigma_j} u_i(s)\).
That is, each player's strategy is a probability measure over its original strategy set (\emph{i.e.}, a mixed strategy), and its utility is the resulting expected utility in the original game.
A mixed-strategy NE is an NE of the mixed-strategy game.%
\footnote{
More generally, one can consider settings where randomness is a limited resource (\emph{e.g.}, only a limited number of random bits are available to the agent) or where the agent can only mix between a limited number of pure strategies (\emph{i.e.}, its mixed strategy must be \emph{sparse}).
Alternatively, its mixed strategy may be restricted to some class of \emph{representable} distributions, such as an explicit parametric model or implicit density model like a Generative Adversarial Network \citep{Goodfellow_2014,goodfellow2020generative}.
}

\paragraph{Continuous game.}
A continuous-action game is a game whose strategy sets are subsets of Euclidean space, \emph{e.g.}, \(\mathcal{S}_i \subseteq \mathbb{R}^d\).
The following theorems apply to such games.
\citet{Nash50:Equilibrium} showed that if each \(\mathcal{S}_i\) is nonempty and finite, a mixed-strategy NE exists.
\citet{Glicksberg52:Further} showed that if each \(\mathcal{S}_i\) is nonempty and compact, and each \(u_i\) is continuous, a mixed-strategy NE exists.
\citet{Glicksberg52:Further,Fan_1952,Debreu_1952} showed that if each \(\mathcal{S}_i\) is nonempty, compact, and convex, and each \(u_i\) is continuous and quasiconcave in \(s_i\), a pure strategy NE exists.
\citet{Dasgupta86:Existence} showed that if each \(\mathcal{S}_i\) is nonempty, compact, convex, and \(u_i\) is upper semicontinuous and graph continuous and quasiconcave in \(s_i\), a pure strategy NE exists.
They also showed that if each \(\mathcal{S}_i\) is nonempty, compact, and convex, and \(u_i\) is bounded and continuous except on a subset (defined by technical conditions) and weakly lower semicontinuous in \(s_i\), and \(\sum_{i \in \mathcal{I}} u_i\) is upper semicontinuous, a mixed-strategy NE exists.
\citet{Rosen_1965} proved the uniqueness of a pure NE for continuous-action games under diagonal strict concavity assumptions.
Most equilibrium-finding algorithms in the literature target discrete-action games, raising the question of how to compute equilibria for continuous-action games.

%% file: method.tex
\section{Method}
\label{sec:method}

In order to build up to our method, we first introduce some key concepts in a gradual fashion.

\paragraph{Pseudo-gradient.}
Let \(d \in \mathbb{N}\) be a natural number,
\(f : \mathbb{R}^d \to \mathbb{R}\) be a function,
\(\mu = \mathcal{N}(\mathbf{0}_d, \mathbf{I}_d)\) be the \(d\)-dimensional standard normal distribution,
\(\sigma \in \mathbb{R}\) be a scalar,
and
\begin{align}
    f_\sigma(\mathbf{x}) = \expect_{\mathbf{z} \sim \mu} f(\mathbf{x} + \sigma \mathbf{z})
\end{align}
Assume \(\sigma \neq 0\).
Then \(f_\sigma = f \ast G_\sigma\), where \(G_\sigma\) is a Gaussian kernel.
It is a property of convolutions that \(\nabla (f \ast G_\sigma) = f \ast \nabla G_\sigma\).
Therefore, since \(G_\sigma\) is smooth (\emph{i.e.}, infinitely differentiable), \(f_\sigma\) is also smooth.
This holds even if \(f\) itself is not even continuous.
In the literature, \(\nabla f_\sigma\) is called the \emph{pseudo-gradient} of \(f\).
It satisfies the identity
\begin{align}
    \nabla f_\sigma(\mathbf{x}) = \expect_{\mathbf{z} \sim \mu} \tfrac{1}{\sigma} f(\mathbf{x} + \sigma \mathbf{z}) \mathbf{z}
\end{align}
This gives us an unbiased estimator that forms the basis of OpenAI ES \citep{Salimans_2017}, a natural evolution strategies method, as described in \S\ref{sec:related_research}.

\paragraph{Applications.}
The pseudo-gradient can be used to perform zeroth-order, \emph{i.e.} gradient-free, optimization of \(f\).
This can be useful in cases where straightforward stochastic gradient ascent is not possible or desirable.
For example, \(f\) might not be differentiable.
Alternatively, it might have gradients for which unbiased estimators are unavailable or have high variance.
The use of pseudo-gradients for optimization has been studied by \citet{Duchi_2015, Nesterov_2017, Shamir_2017, Salimans_2017, Berahas_2022, metz2021gradients}, among others.
It has been shown to be a scalable alternative to classical methods in reinforcement learning \citep{Salimans_2017}.

\paragraph{Variance reduction.}
We define
\begin{align}
    \mathbf{g}_\text{SP} &= \tfrac{1}{\sigma} f(\mathbf{x} + \sigma \mathbf{z}) \mathbf{z} \\
    \mathbf{g}_\text{FD} &= \tfrac{1}{\sigma} (f(\mathbf{x} + \sigma \mathbf{z}) - f(\mathbf{x})) \mathbf{z} \\
    \mathbf{g}_\text{CD} &= \tfrac{1}{2 \sigma} (f(\mathbf{x} + \sigma \mathbf{z}) - f(\mathbf{x} - \sigma \mathbf{z})) \mathbf{z}
\end{align}
as the \emph{single-point (SP)}, \emph{forward-difference (FD)}, and \emph{centered-difference (CD)} (or antithetic) estimators, respectively.
These are all unbiased estimators of the pseudo-gradient.
However, the first has a large variance, so the latter two are typically used in practice \cite{Berahas_2022}.
The variance of the pseudo-gradient estimator can be further reduced by taking a mean over many perturbations, as in \(\tfrac{1}{2 n \sigma} \sum_{i=1}^n (f(\mathbf{x} + \sigma \mathbf{z}_i) - f(\mathbf{x} - \sigma \mathbf{z}_i)) \mathbf{z}_i\), where \(\mathbf{z}_i \sim \mu\) are independent.
Other variance reduction techniques are surveyed in \citet{mohamed2020monte}, among other works.

\paragraph{Pseudo-Jacobian.}
We extend the preceding concept of the pseudo-gradient from a scalar-valued function to a vector-valued function.
Let \(n \in \mathbb{N}\),
\(\mathbf{f} : \mathbb{R}^d \to \mathbb{R}^n\),
and
\begin{align}
    \mathbf{f}_\sigma(\mathbf{x}) = \expect_{\mathbf{z} \sim \mu} \mathbf{f}(\mathbf{x} + \sigma \mathbf{z})
\end{align}
By analogy with the pseudo-gradient, we call \(\nabla \mathbf{f}_\sigma\) the \emph{pseudo-Jacobian} of \(\mathbf{f}\).
Furthermore, it satisfies the identity
\begin{align}
    \nabla \mathbf{f}_\sigma(\mathbf{x}) = \expect_{\mathbf{z} \sim \mu} \tfrac{1}{\sigma} \mathbf{f}(\mathbf{x} + \sigma \mathbf{z}) \otimes \mathbf{z}
\end{align}
Therefore, we have an unbiased estimator for it.

\paragraph{Simultaneous gradient.}
One common approach to game-solving in the literature is \emph{simultaneous gradient ascent (SGA)}, which is defined as follows.
Let \(\mathbf{u} : \mathbb{R}^{n \times d} \to \mathbb{R}^n\) be a utility function, where \(n\) is the number of players and \(d\) is the dimensionality of each player's strategy parameters.
The \emph{simultaneous gradient} of \(\mathbf{u}\) is the function \(\mathbf{v} : \mathbb{R}^{n \times d} \to \mathbb{R}^{n \times d}\) where \(\mathbf{v}_i = \nabla_i \mathbf{u}_i\).
That is, it is the derivative of each parameter with respect to utility of the player that parameter belongs to.
Equivalently, \(\mathbf{v} = \diag \nabla \mathbf{u}\), where \(\nabla \mathbf{u}\) is the Jacobian of \(\mathbf{u}\).
SGA consists of running the ODE \(\frac{\mathrm{d}}{\mathrm{d}t} \mathbf{x} = \mathbf{v}(\mathbf{x})\).
That is, each player tries to greedily increase their own utility, acting as if the other players are fixed.
Explicitly, it uses the iteration scheme \(\mathbf{x}_{t+1} = \mathbf{x}_t + \alpha_t \mathbf{v}(\mathbf{x}_t)\) for \(t \in \mathbb{N}\), where \(\alpha_t \in \mathbb{R}\) is a stepsize.

\citet{Mertikopoulos_2019} analyze the conditions under which SGA converges to Nash equilibria.
They prove that, if the game admits a pseudoconcave potential or if it is monotone, the players' actions converge to Nash equilibrium, no matter the level of uncertainty affecting the players' feedback.
\citet{bichler2021learning} write that most auctions in the literature assume symmetric bidders and symmetric equilibrium bid functions~\cite{Krishna02:Auction}.
This symmetry creates a potential game, and SGA provably converges to a pure local Nash equilibria in finite-dimensional continuous potential games \citep{Mazumdar_2020}.
Thus in any symmetric and smooth auction game, symmetric gradient ascent with appropriate (square-summable but not summable) step sizes almost surely converges to a local ex-ante approximate Bayes-Nash equilibrium~\cite[Proposition 1]{bichler2021learning}.

\paragraph{Optimistic gradient.}
Another approach to game-solving in the literature is \emph{optimistic gradient ascent (OGA)}, which iterates \(\mathbf{x}_{t+1} = \mathbf{x}_t + \alpha_t \mathbf{v}(\mathbf{x}_t) + \beta_t (\mathbf{v}(\mathbf{x}_t) - \mathbf{v}(\mathbf{x}_{t-1}))\) for \(t \in \mathbb{N}\), where \(\alpha_t, \beta_t \in \mathbb{R}\).
In the standard version of OGA, \(\alpha_t = \beta_t\).
OGA uses the past simultaneous gradient \(\mathbf{v}(\mathbf{x}_{t-1})\) to create an extrapolation or prediction of the future simultaneous gradient \(\mathbf{v}(\mathbf{x}_{t+1})\), and updates according to this prediction.
OGA converges in some games where SGA fails to converge or diverges.
OGA has been analyzed by \citet{Popov1980}, \citet{Daskalakis_2017}, and \citet{Hsieh_2019}, among others.\footnote{There are also other learning dynamics in the literature, which are surveyed and analyzed by \citet{Balduzzi_2018}, \citet{Letcher_2018}, \citet{Letcher_2019}, \citet{Mertikopoulos_2019}, \citet{Mazumdar_2019}, \citet{Hsieh_2021}, and \citet{Willi_2022}, among others, but these are beyond the scope of this paper.}

\paragraph{Simultaneous pseudo-gradient.}
Both SGA and OGA, as well as other game-solving approaches in the literature, require computing the simultaneous gradient \(\mathbf{v}\).
However, in some situations, \(\mathbf{v}\) does not exist because \(\mathbf{u}\) is not differentiable.
In other situations, \(\mathbf{u}\) \emph{is} differentiable, but obtaining an unbiased estimator of its gradient is difficult or intractable.
This can happen if, for example, \(\mathbf{u}\) is an expectation (with respect to a distribution parameterized by \(\mathbf{x}\)) of some non-differentiable function.
An example of such a situation is an auction, which we will describe in \S\ref{sec:experiments}.

To resolve this problem, we replace the gradient of \(\mathbf{u}_i\) in the definition of \(\mathbf{v}\) with a pseudo-gradient.
Explicitly,
\begin{align}
    \mathbf{g}_i = \tfrac{1}{\sigma} \mathbf{u}(\mathbf{x}_i + \sigma \mathbf{z}_i, \mathbf{x}_{-i})_i \mathbf{z}_i
\end{align}
for each Player \(i\), where \(\mathbf{z}_i \sim \mu_i\) and \(\mu_i\) is a multivariate standard normal distribution of the same dimension as \(\mathbf{x}_i\).
In other words, we estimate the pseudo-gradient of \(\mathbf{u}_i\) (which is a scalar-valued function, since it outputs only the utility of Player \(i\)) with respect to the parameters of Player \(i\).
This is the approach taken by \citet{bichler2021learning}.
It requires one perturbation for each player and subsequent evaluation of \(\mathbf{u}\).
Therefore, the number of utility function evaluations per iteration is linear in the number of players.

\paragraph{Joint perturbation.}
Inspired by all of the preceding concepts that have been discussed, we combine
(1) the identity \(\mathbf{v} = \diag \nabla \mathbf{u}\),
(2) the concept of the pseudo-Jacobian,
and (3) the identity \(\diag (\mathbf{a} \otimes \mathbf{b}) = \mathbf{a} \odot \mathbf{b}\) to obtain an estimator that requires only a \emph{single}, joint perturbation across all players.
In particular, we notice that
\begin{align}
    \mathbf{v}_\sigma(\mathbf{x})
    &= (\mathbf{v} \ast G_\sigma)(\mathbf{x}) \\
    &= ((\diag \nabla \mathbf{u}) \ast G_\sigma)(\mathbf{x}) \\
    &= \diag (\nabla \mathbf{u} \ast G_\sigma) (\mathbf{x}) \\
    &= \diag \nabla (\mathbf{u} \ast G_\sigma) (\mathbf{x}) \\
    &= \diag \nabla \mathbf{u}_\sigma (\mathbf{x}) \\
    &= \diag \expect_{\mathbf{z} \sim \mu} \tfrac{1}{\sigma} \mathbf{u}(\mathbf{x} + \sigma \mathbf{z}) \otimes \mathbf{z} \\
    &= \expect_{\mathbf{z} \sim \mu} \tfrac{1}{\sigma} \diag \mathbf{u}(\mathbf{x} + \sigma \mathbf{z}) \otimes \mathbf{z} \\
    &= \expect_{\mathbf{z} \sim \mu} \tfrac{1}{\sigma} \mathbf{u}(\mathbf{x} + \sigma \mathbf{z}) \odot \mathbf{z}
\end{align}
Therefore, we obtain the following unbiased estimator:
\begin{align}
    \mathbf{g} = \tfrac{1}{\sigma} \mathbf{u}(\mathbf{x} + \sigma \mathbf{z}) \odot \mathbf{z}
\end{align}
In terms of indices, for clarity:
\begin{align}
    \mathbf{g}_i = \tfrac{1}{\sigma} \mathbf{u}(\mathbf{x} + \sigma \mathbf{z})_i \mathbf{z}_i
\end{align}
With this new estimator, the number of utility function evaluations per iteration is now \emph{constant} in the number of players, rather than linear.
This dramatically reduces the number of utility function evaluations when there are many players.
Therefore, it makes game solving significantly more efficient in many-player games, especially when the utility function is expensive to evaluate in terms of wall time, memory, money, or other resources such as real-world experiments.
We can use this simultaneous pseudo-gradient estimator inside schemes like simultaneous gradient ascent, extragradient ascent \citep{korpelevich_1976}, optimistic gradient ascent \citep{Popov1980,Daskalakis_2017}, and so on.
We call our method \emph{joint-perturbation simultaneous pseudo-gradient (JPSPG)}, in contrast to the classical method, \emph{simultaneous pseudo-gradient (SPG)}.

%% file: experiments/main.tex
\section{Experiments}
\label{sec:experiments}

We test our approach against the classical approach on several continuous-action games, in particular on many kinds of auction and on continuous Goofspiel (which can be thought of as a kind of auction with budget constraints). 
Our experimental hyperparameters are as follows.
For each experiment, we run 8 trials.
In each graph, solid lines show the mean across trials, and bands show the standard error of the mean.
The classical method is shown in blue, while our method is shown in orange.
We use a stepsize of \(10^{-4}\).
For the Gaussian smoothing, we use a perturbation scale \(\sigma\) of \(10^{-1}\).
We use a batch size per iteration of \(256\).
To update parameters, we use the AdaBelief optimizer \citep{zhuang2020adabelief}.
For each player's strategy network, we use a single hidden layer of size 64, the ReLU activation function, and He initialization \citep{he_2015} for initializing the network's weights.
Each experiment was run individually on one NVIDIA A100 SXM4 40GB GPU.

We estimate the exploitability of the learned strategy profile as follows. First, we compute approximate best responses for each player via reinforcement learning, specifically OpenAI ES \citep{Salimans_2017}.
For this, we use the same batch size and optimizer as before.
The best responses are trained for \(1024\) iterations.
Second, as described in Section~\ref{sec:formulation}, exploitability is defined as \(\Phi(s) = \sum_{i \in \mathcal{I}} (u_i(b_i, s_{-i}) - u_i(s))\), where \(b_i\) is the best-response strategy for player \(i\) to \(s\).
We estimate each occurrence of \(u_i\) in this expression by averaging over \(1024\) samples of game play.

Next we proceed to presenting the benchmark settings and the performance of the methods on those settings.

\input{experiments/unit_demand_auction}

\input{experiments/knapsack_auction}

\input{experiments/sequential_auction}

\input{experiments/goofspiel}

%% file: experiments/unit_demand_auction.tex
\subsection{Multi-item unit-demand auction}

An auction is a mechanism by which a set of \emph{items} are sold to a set of \emph{bidders}, who have valuations for items or sets thereof.
Auctions play a central role in the study of markets and are used in a wide range of real-world contexts \citep{Krishna02:Auction}, such as advertising, commodities, radio spectrum allocation, real estate, and more.
To evaluate their method, \citet{bichler2021learning} used auctions as a benchmark.

Here, we consider a type of multi-item auction called a \emph{unit-demand auction}.
In this auction, we have \(n\) bidders and \(m\) non-identical items. Each bidder \(i\) has a private valuation \(v_{ij}\) for each item \(j\).
Furthermore, each bidder has \emph{unit demand}, meaning that its value for a \emph{bundle} of items is the same as that for the maximum-value item in that bundle: \(v_i(S) = \max_{j \in S} v_{ij}\), where \(S\) is a bundle of items.
Housing markets are often given as an example of unit-demand preferences.
This model was first studied by \citet{shapley1971assignment}.
This is a special case of a \emph{limited-demand model} with \(K\) units in which each bidder has use for at most \(L < K\) units, as described in \citet[\S13.4.2 and \S13.5.2]{Krishna02:Auction}.
The single-unit case corresponds to \(L = 1\).

For our experiment, we use a prior where bidder-item valuations \(v_{ij}\) are independently sampled uniformly at random from the unit interval.
Each player \(i\) submits a bid \(b_{ij}\) for each item \(j\).
To allocate items, our auction mechanism assigns items to bidders in a way that maximizes the sum of bids across players.
This requires solving a \emph{linear assignment problem}, which can be described as follows.
Given a bid matrix \(b \in \mathbb{R}^{n \times m}\), compute a binary assignment \(x \in \{0, 1\}^{n \times m}\) that satisfies the following:
\begin{align}
    \text{maximize} \quad & \sum_{i \in [n]} \sum_{j \in [m]} b_{ij} x_{ij} \\
    \text{subject to} \quad
    & \sum_{i \in [n]} x_{ij} \leq 1 & \forall j \in [m] \\
    & \sum_{j \in [m]} x_{ij} \leq 1 & \forall i \in [n] \\
    & x_{ij} \in \{0, 1\} & \forall i \in [n], j \in [m]
\end{align}
That is, maximize the sum of values subject to the constraint that each item is assigned to at most one bidder and each bidder receives at most one item.

The linear assignment problem was first described in a seminal paper by \citet{kuhn1955hungarian}, who introduced a solution approach called the Hungarian method.
Subsequently, various algorithms have been devised in the literature.
We use the modified Jonker-Volgenant algorithm \citep{jonker1988shortest} given in \citet{crouse2016implementing}, as implemented in the Python scientific computing library scipy \citep{2020SciPy-NMeth}, and reimplement it in JAX \citep{jax2018github}.
This algorithm has a time complexity of \(O(n^3)\).

\begin{figure}[!ht]
    \centering
    \includegraphics{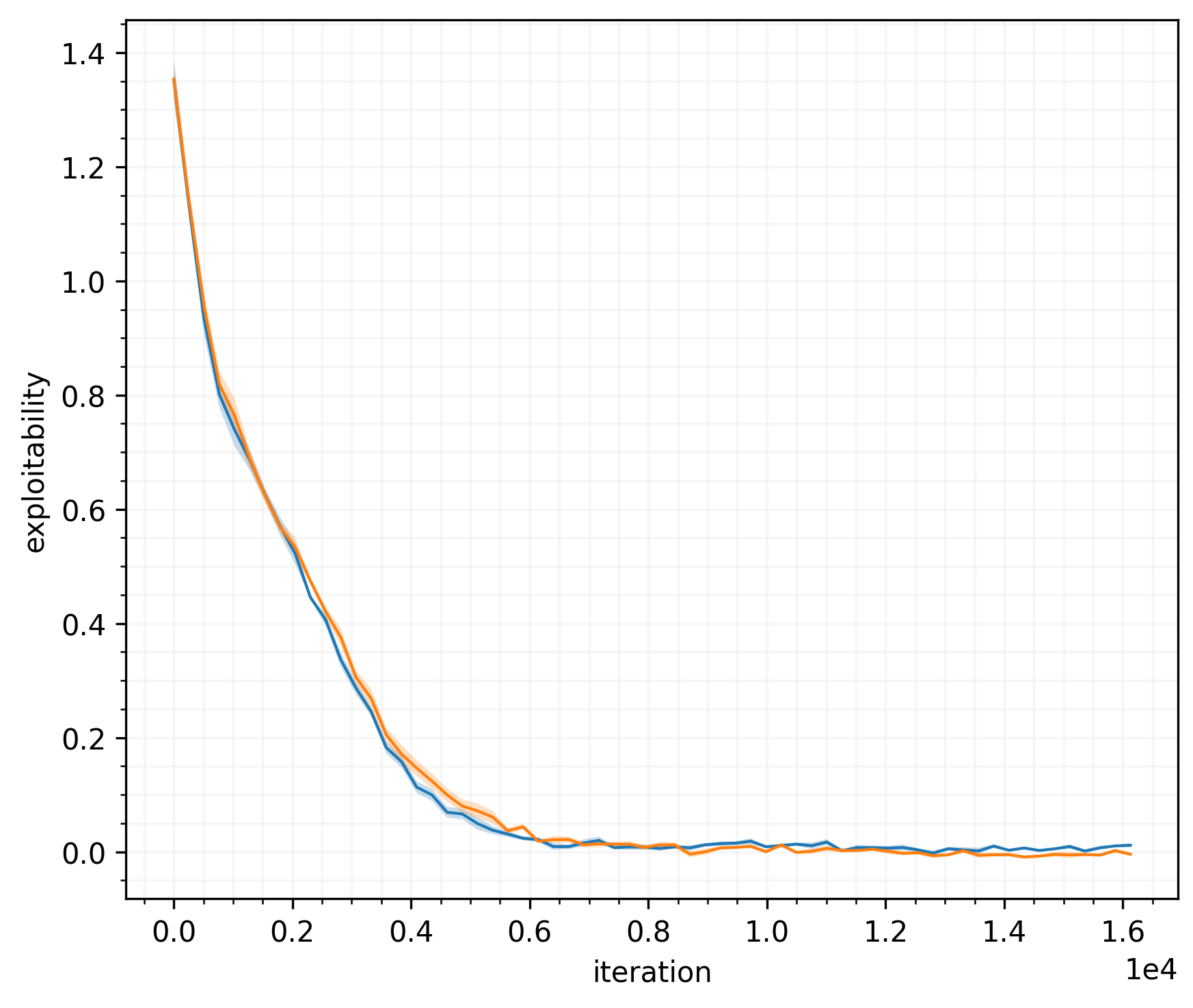}
    \includegraphics{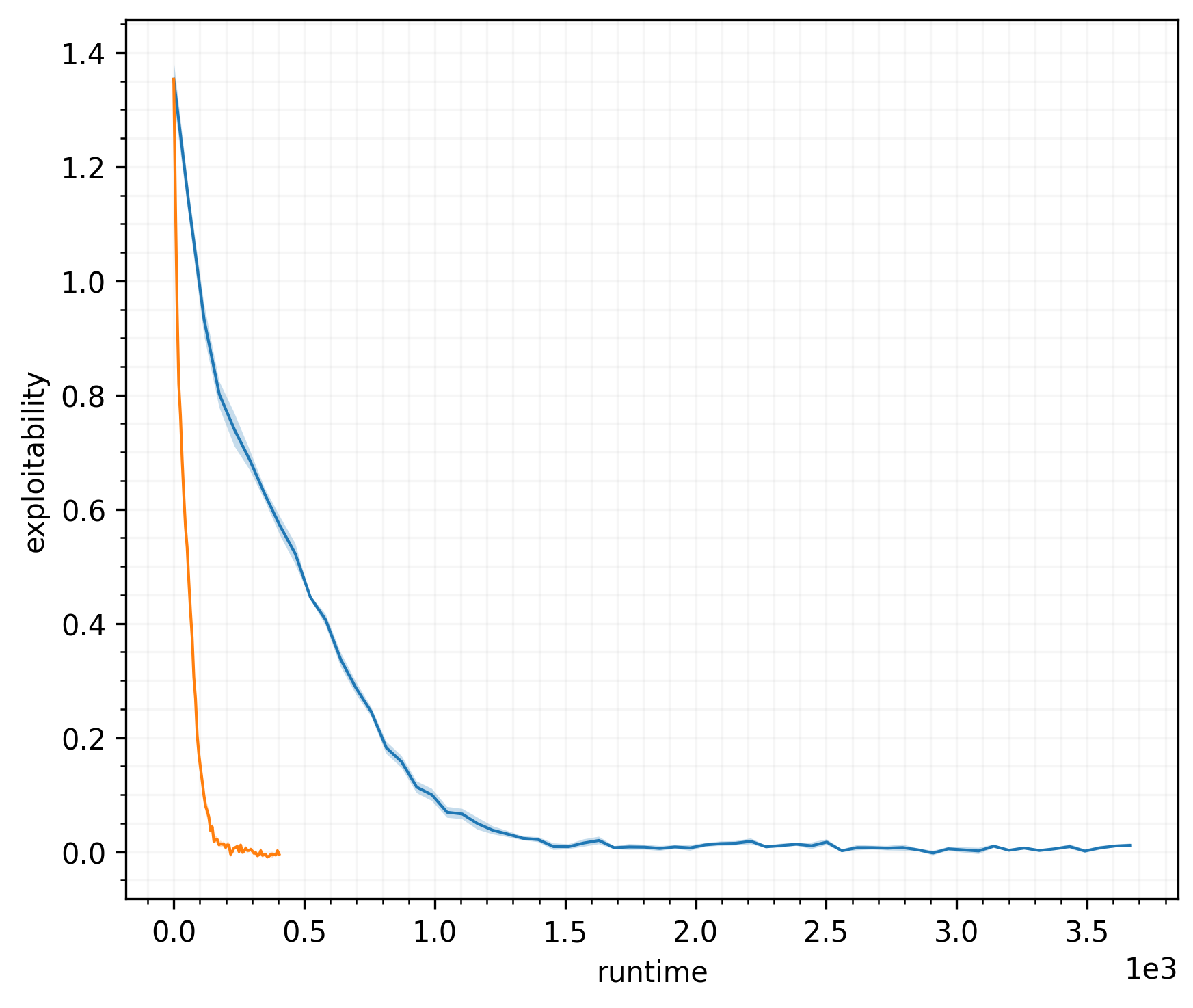}
    \caption{10-player, 10-item unit-demand auction.}
    \label{fig:unit_demand_auction}
\end{figure}

\begin{figure}[!ht]
    \centering
    \includegraphics{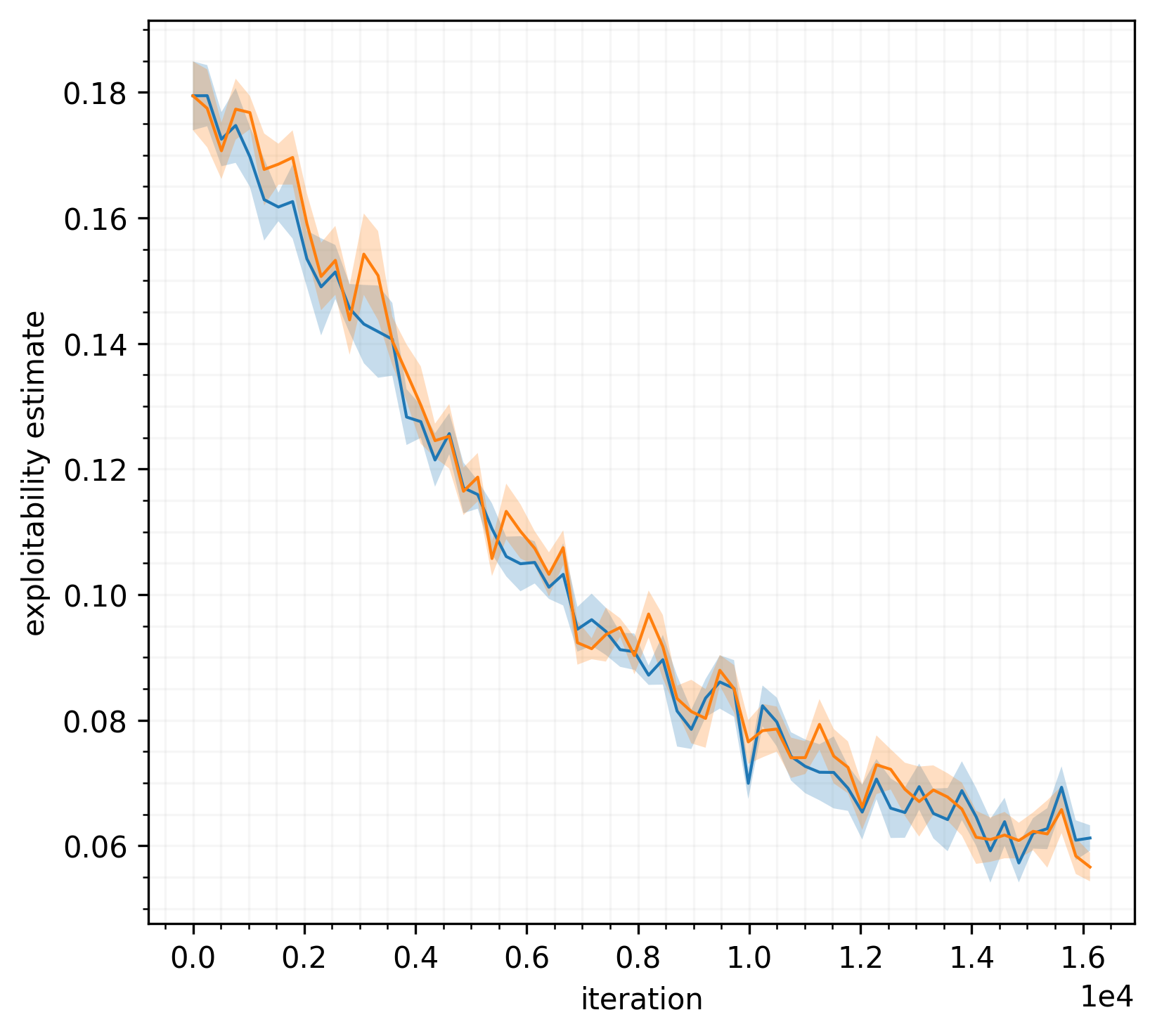}
    \includegraphics{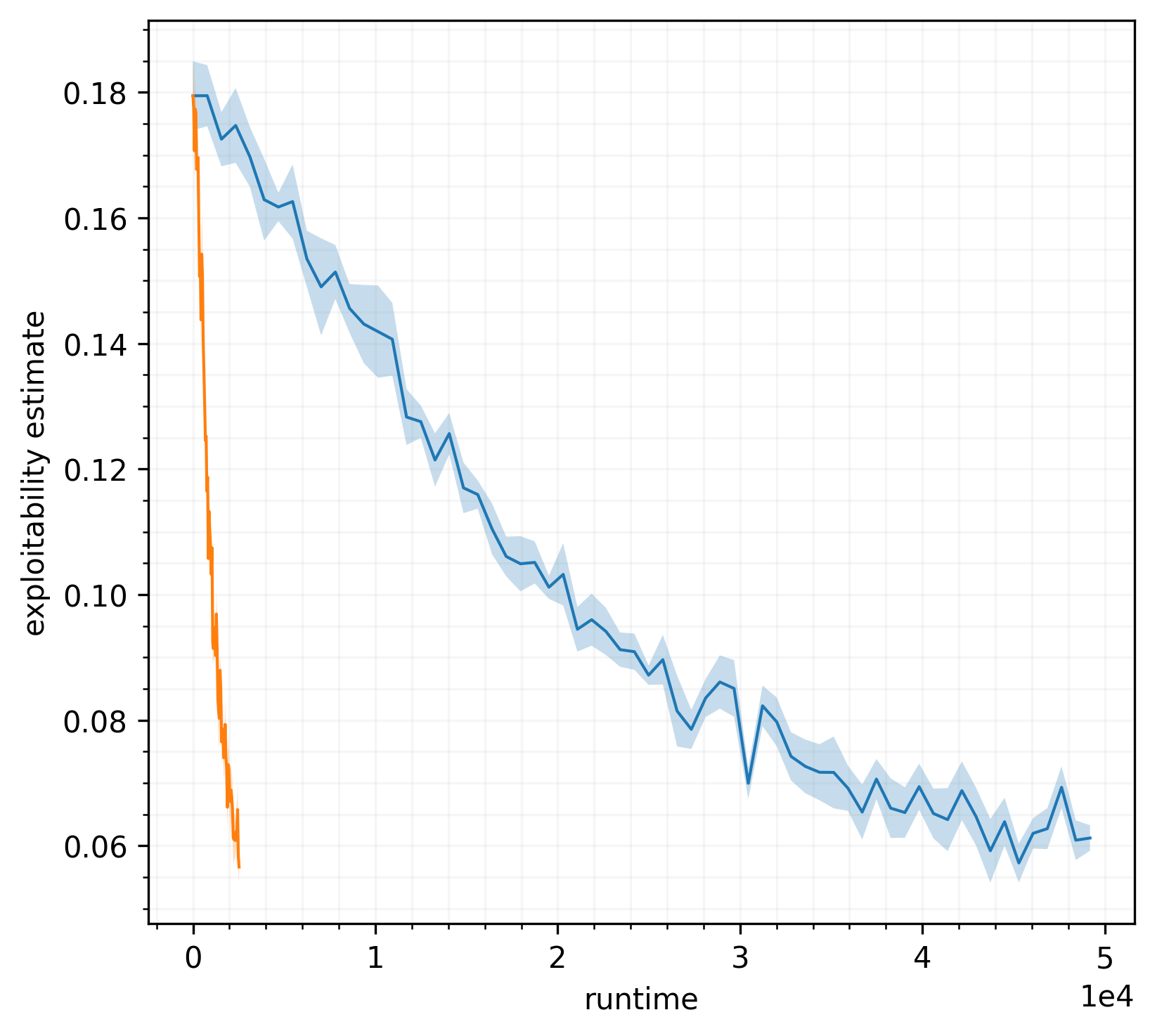}
    \caption{20-player, 20-item unit-demand auction.}
    \label{fig:unit_demand_auction_20p}
\end{figure}

Figures~\ref{fig:unit_demand_auction} and~\ref{fig:unit_demand_auction_20p} show the exploitability over the course of training for a 10-player, 10-item unit-demand auction and 20-player, 20-item unit-demand auction, respectively.
The figures shows that both our method and the classical method attain a similar exploitability for a given iteration count, but our method is significantly faster in terms of run time (here expressed in seconds).
This is to be expected, because the standard method requires $n \in \{10,20\}$ evaluations of the utility function per batch instance, each of which requires solving an assignment problem, which is expensive, whereas ours only requires a \emph{single} utility function evaluation. The advantage of our method increases as the size of the problem instance grows. 

%% file: experiments/knapsack_auction.tex
\subsection{Knapsack auction}

Another type of auction is a \emph{knapsack auction}.
We follow the description given in \citet{aggarwal2006knapsack}.
In a knapsack auction, an auctioneer auctions off space in a knapsack of known capacity \(C\).
Each player seeks to place exactly one object in the knapsack.
Player \(i\) values the placement of its object in the knapsack at \(v_i\).
The valuations are private data of each respective player.
Each object takes up a certain amount of space in the knapsack.
Player \(i\)'s object takes space \(c_i\), and these sizes are publicly known.
Thus, the \(c_i\)s are public while the \(v_i\)s are private.
Each player submits a bid \(b_i\).
Knapsack auctions are also studied in \citet{dutting2014performance} and \citet{berg2010knapsack}, who model the problem of bidding in ad auctions as a penalized multiple choice knapsack problem.

As \citet{aggarwal2006knapsack} note, the knapsack auction problem models several interesting applications.
For example, consider running a single auction to sell advertising space on a web page over the course of a day.
Suppose statistical information is available for each advertiser as to how many showings (\emph{i.e.}, impressions) are necessary for to result in a user click-through and as well how many times the web page itself will be viewed in a day.
The number of impressions necessary to generate a click-through corresponds to the \(c_i\)s and the number of total views corresponds to the capacity of the knapsack, \(C\).

Each utility function evaluation requires solving an optimization problem of the form:
\begin{align}
    \text{maximize} \quad & \mathbf{x} \cdot \mathbf{b} \\
    \text{subject to} \quad & \mathbf{x} \cdot \mathbf{c} \leq C \\
    & \mathbf{x} \in \{0, 1\}^n
\end{align}
where \(n\) is the number of players, \(\mathbf{b}\) is the vector of stated values (bids) for each player, \(\mathbf{c}\) is the corresponding vector of sizes, \(\mathbf{x}\) is a binary vector indicating whether each player is included in the knapsack.
Player \(i\)'s final utility is \((v_i - b_i) x_i\).
That is, it is the difference between their private value and their submitted bid, assuming they are included in the knapsack, and zero otherwise.

\begin{figure}[!ht]
    \centering
    \includegraphics{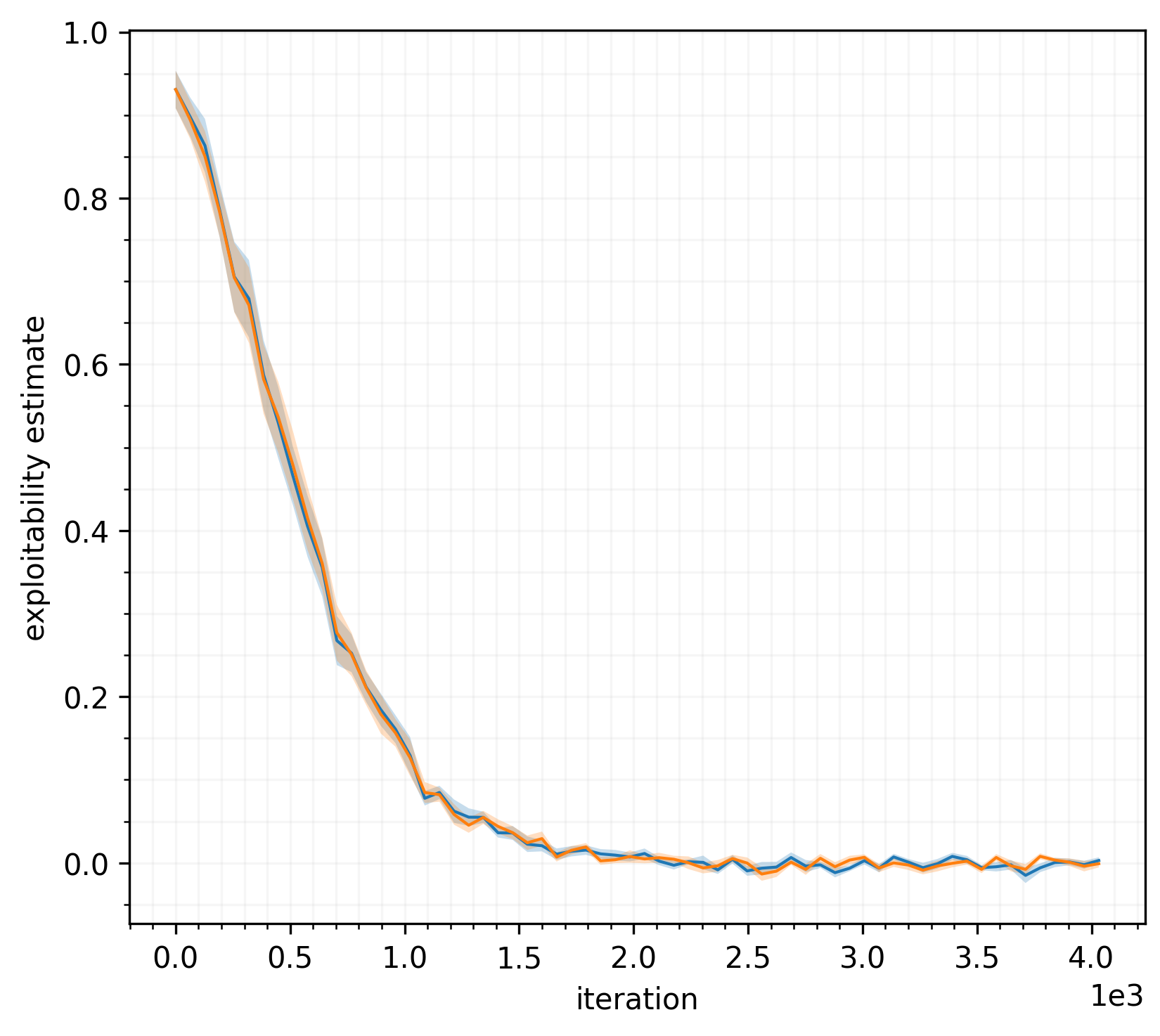}
    \includegraphics{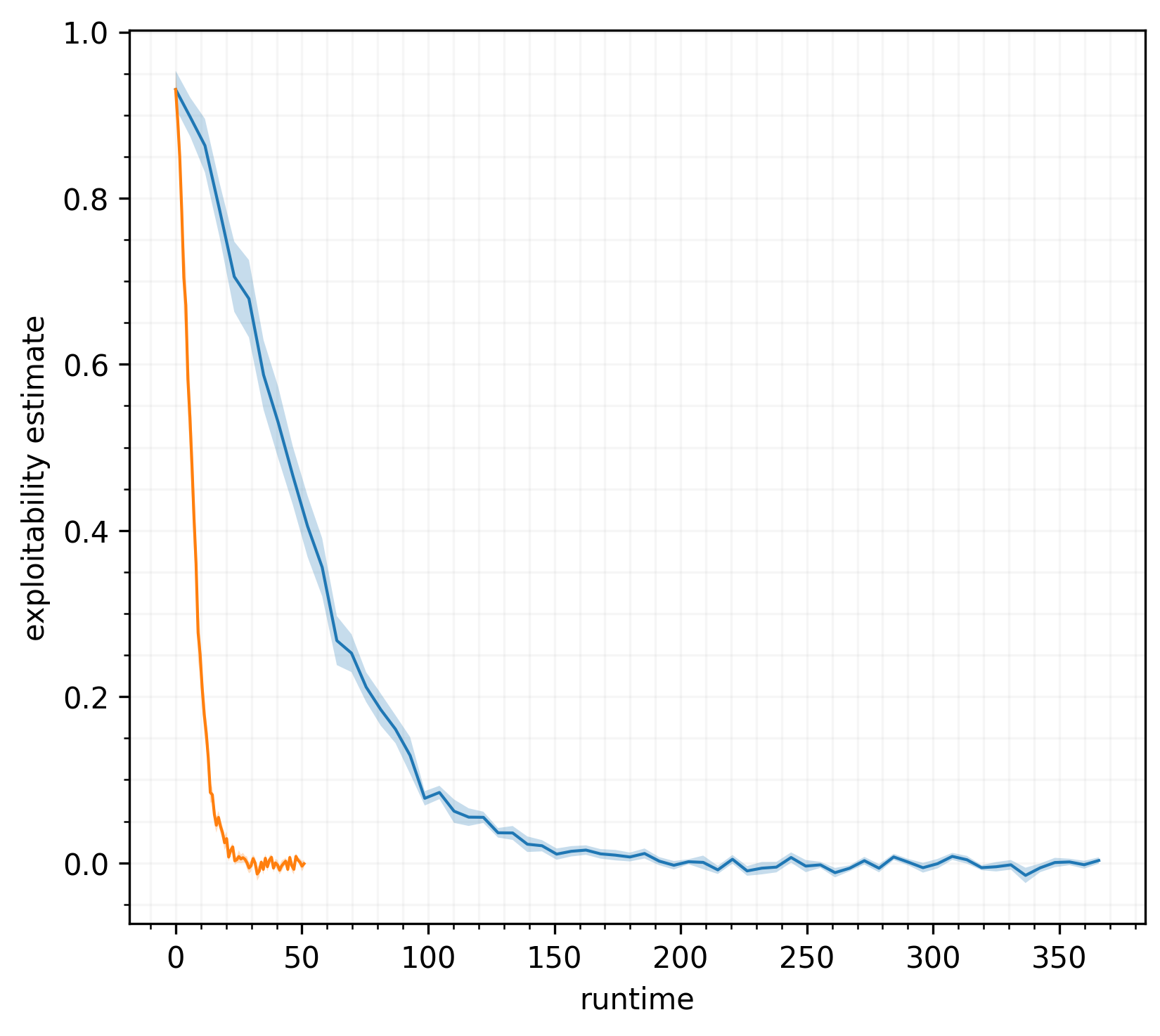}
    \caption{10-player knapsack auction.}
    \label{fig:knapsack_auction_10p}
\end{figure}

\begin{figure}[!ht]
    \centering
    \includegraphics{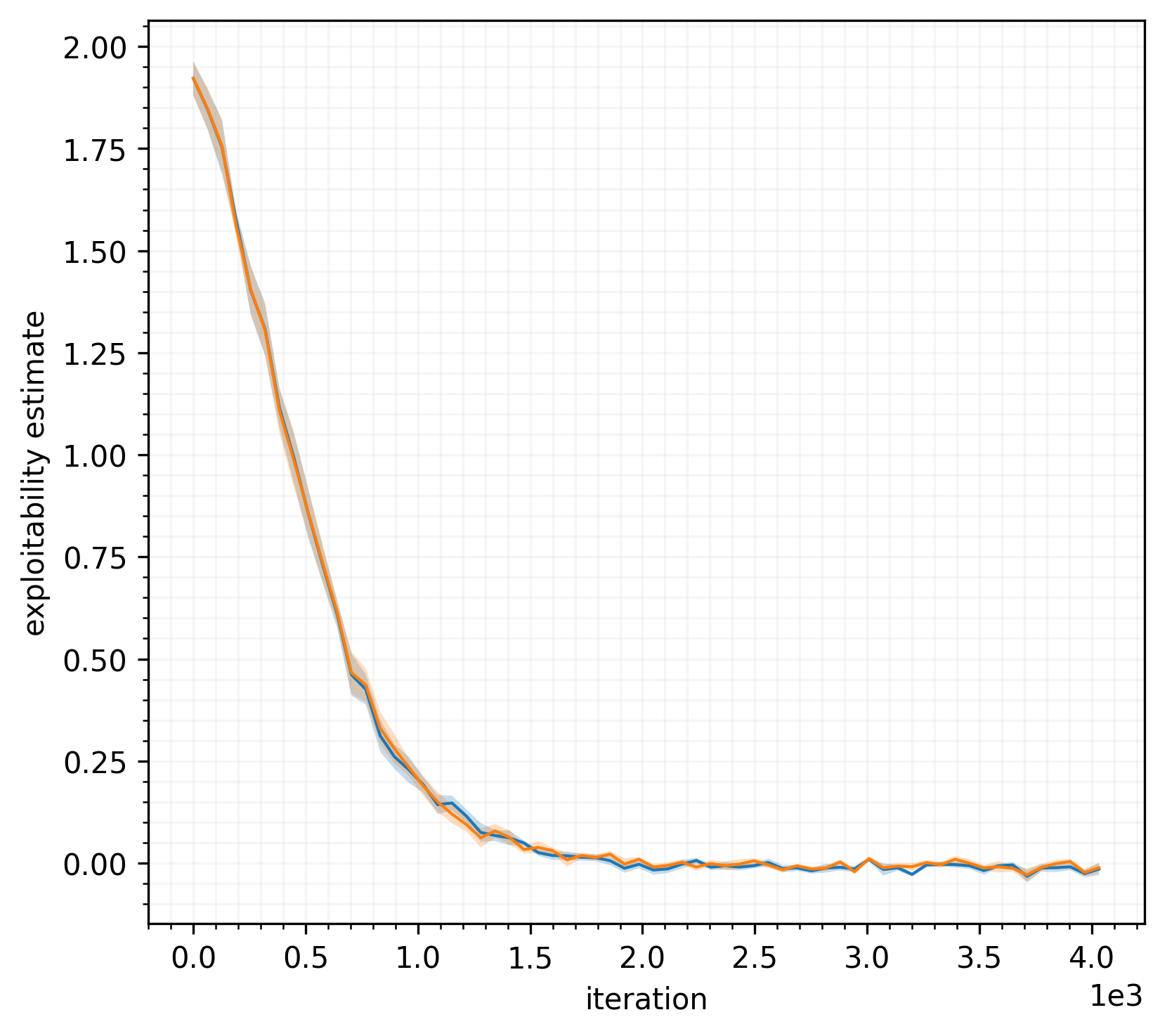}
    \includegraphics{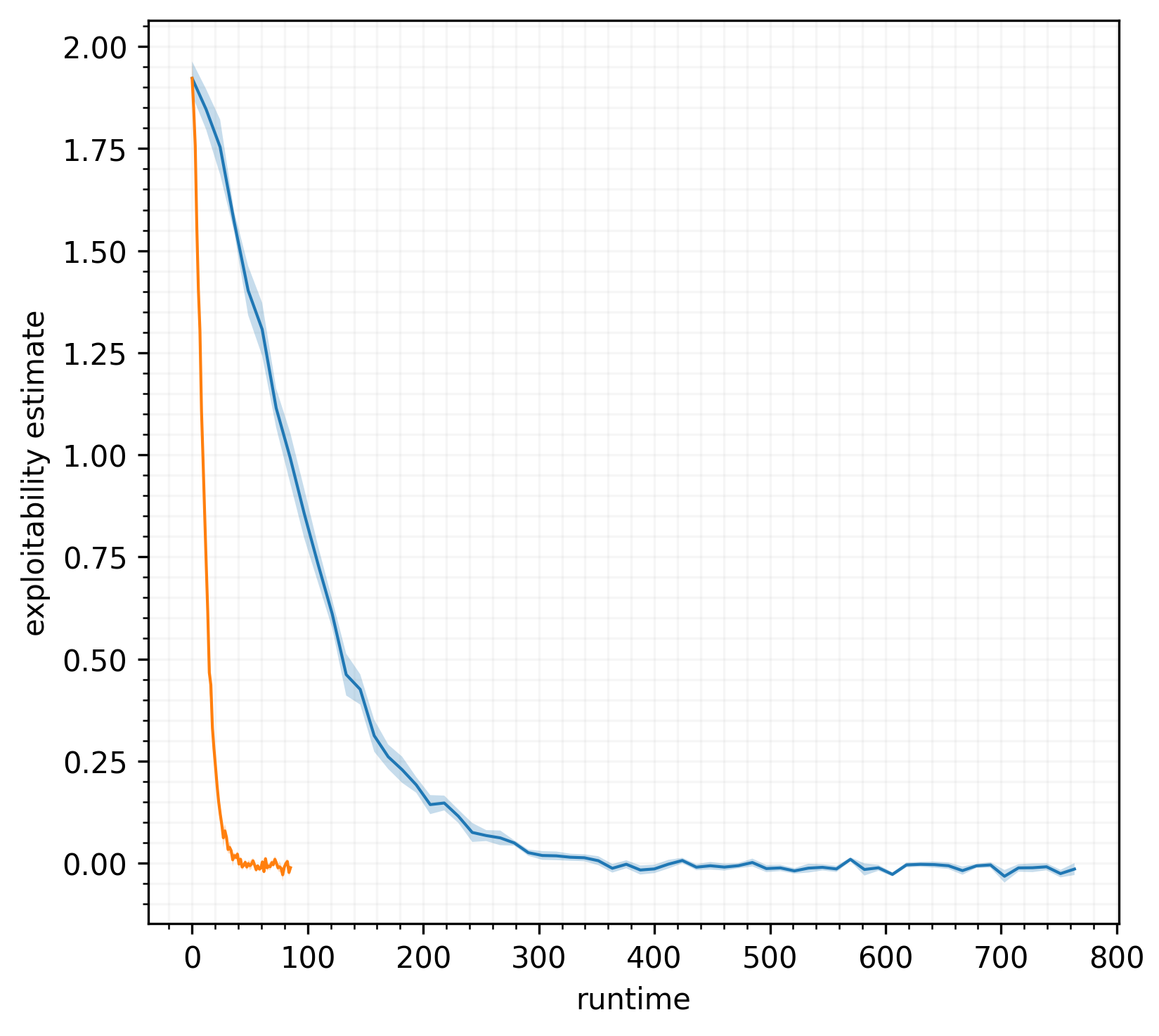}
    \caption{20-player knapsack auction.}
    \label{fig:knapsack_auction_20p}
\end{figure}

This problem can be solved using \emph{integer linear programming (ILP)}.
For this, we use the \texttt{milp} function included in SciPy's library \citep{2020SciPy-NMeth}, which uses the HiGHS optimization solver \citep{huangfu2018parallelizing, hall2023highs}.
Solving an integer program can be expensive (integer linear programming is NP-hard), so reducing the number of utility function evaluations during learning should result in a significant speedup.
In our experiment, we sample the \(v_i\)s and \(c_i\)s from the standard uniform distribution, and sample \(C\) from the standard uniform distribution on \([0, n]\).
Experimental results on the knapsack auction are shown in Figure \ref{fig:knapsack_auction_10p} for 10 players and Figure \ref{fig:knapsack_auction_20p} for 20 players.
As expected, our method requires fewer utility function evaluations per iteration and thus yields a significantly lower training time for attaining the same exploitability.

%% file: experiments/sequential_auction.tex
\subsection{Sequential auction for multiple identical items}

\begin{figure}[!ht]
    \centering
    \includegraphics{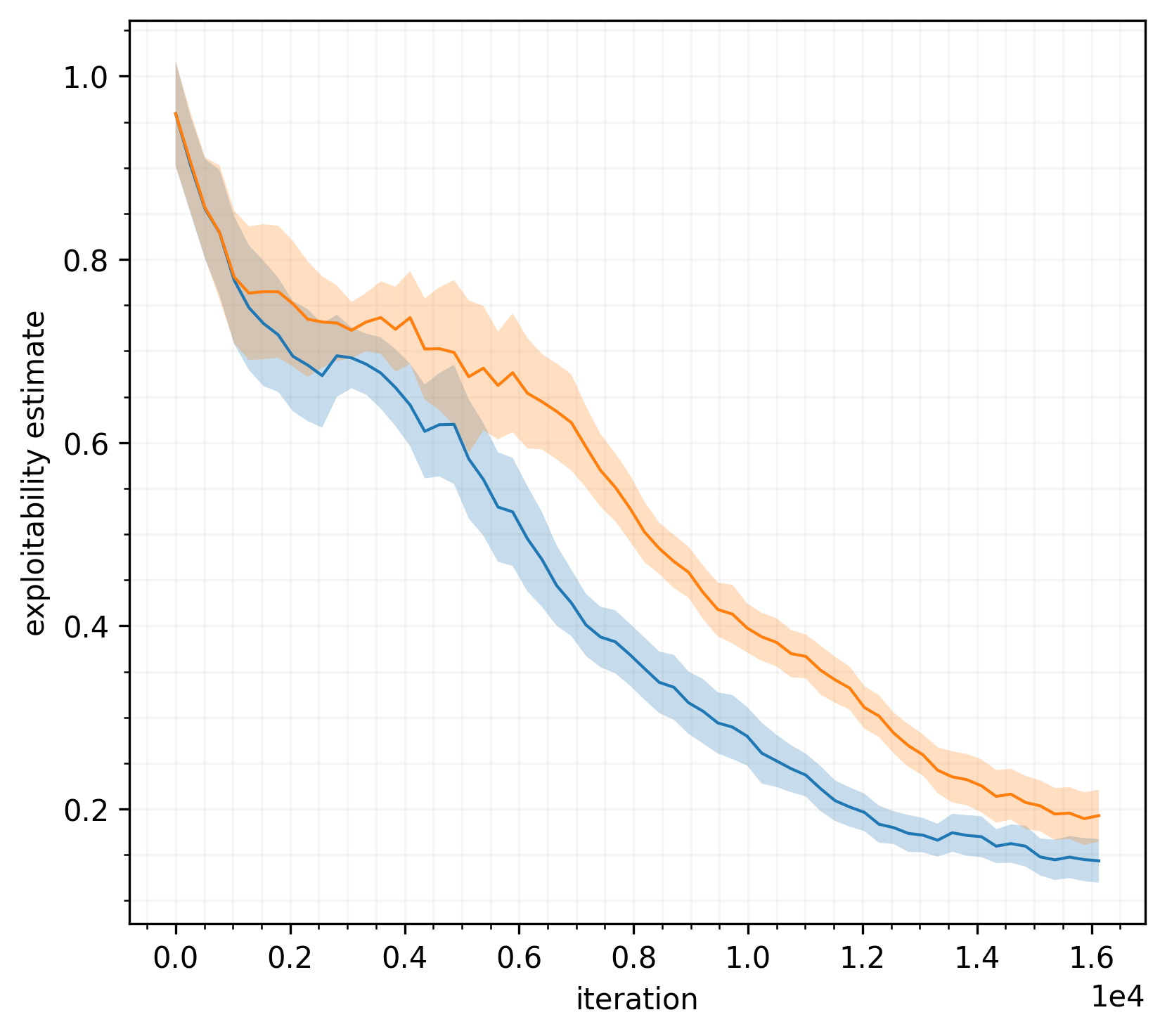}
    \includegraphics{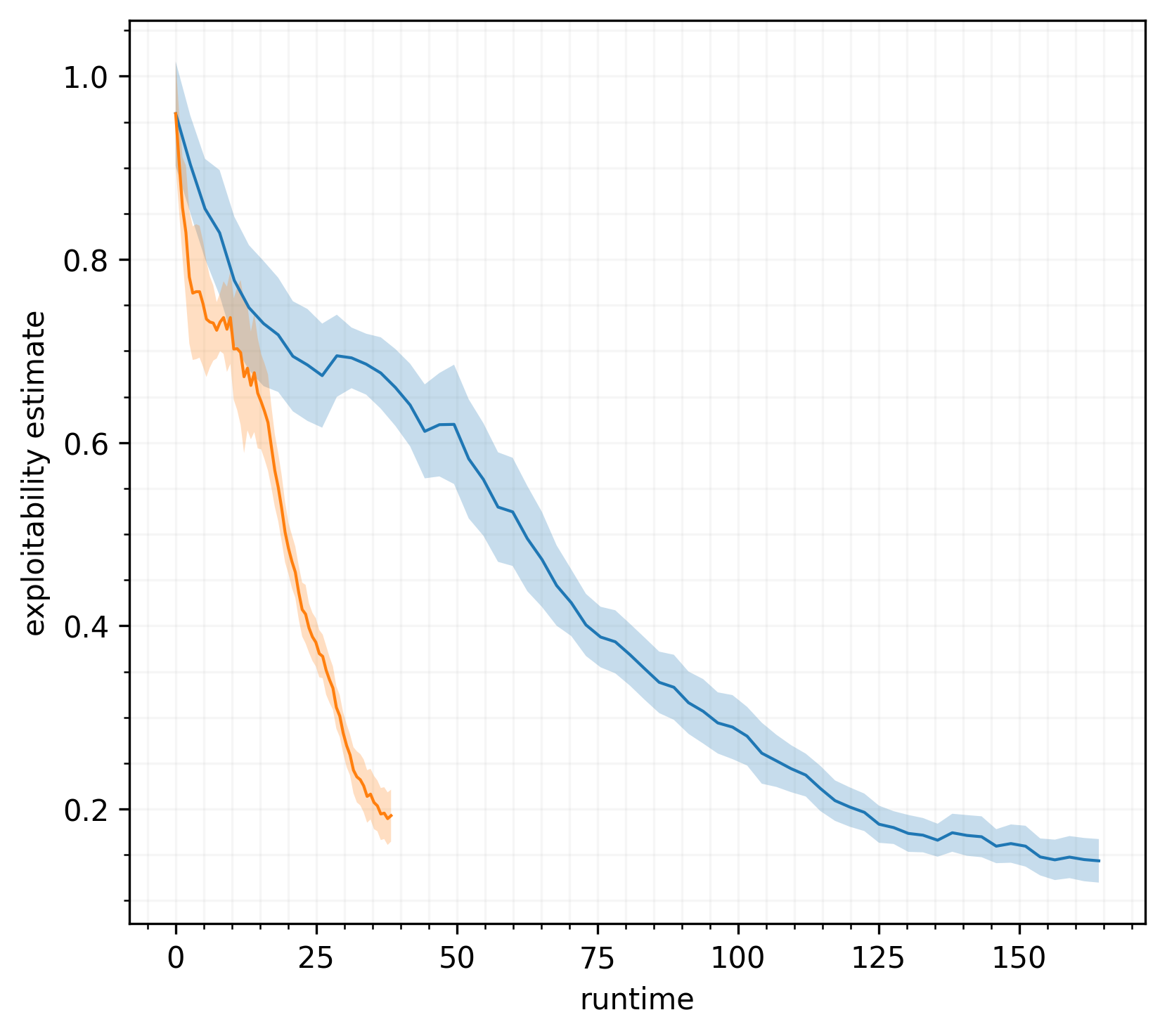}
    \caption{10-player, 5-item sequential auction.}
    \label{fig:sequential_auction}
\end{figure}

Consider a multi-item unit-demand auction in which identical items are sold \emph{sequentially}, rather than simultaneously.
We follow the description given in \citet[\S15.1]{Krishna02:Auction}.
In this auction, \(K\) identical items are sold to \(N > K\) bidders using a sequence of first-price sealed-bid auctions.
Specifically, on each of \(K\) rounds, one of the items is auctioned using the first-price format, and the price at which it is sold---the winning bid---is announced.
We focus on the \emph{single-unit demand} setting, in which each bidder has use for at most one unit.
Thus a bidder leaves the game once it has won an item.
Each bidder has a private value \(v_i\) is that is sampled from the standard uniform distribution.
On round \(K\), a bidder's observation consists of its own private value as well as all the prices of the preceding \(k-1\) rounds, \(p_1, p_2, \ldots, p_{k-1}\).
Results are shown in Figure~\ref{fig:sequential_auction}.
Our method performs much better in terms of wall time.

%% file: experiments/goofspiel.tex
\subsection{Continuous-action Goofspiel}

Goofspiel, also known as \emph{the game of pure strategy}, is a card game invented by mathematician Merrill Flood in the 1930s~\citep{Tucker_1984}.
This game is played with a standard 52-card deck.
The cards of one suit are given to one player, the cards of a second suit are given to the other player, and the cards of a third suit are shuffled and placed face down in the middle.
The cards are valued from low to high as 1 (Ace), 2, 3, \ldots, 10, 11 (Jack), 12 (Queen), and 13 (King).
A round consists of turning up the next card from the middle pile and letting the players simultaneously ``bid'' on this ``prize'' card.
Players bid by choosing one of their own cards and revealing it at the same time as the other player.
The player with the highest bid wins the value of the prize card.
In a tie, the prize value is split between the players.
All three cards are then discarded.
The game ends after 13 rounds, and the winner is the player with the highest score.
Because of its simple mechanics but complex strategy, Goofspiel is commonly used as an example in game theory and artificial intelligence~\citep{Ross71:Goofspiel, Dror_1989, Ferguson_2001, Grimes_2013, Rhoads_2012, Lanctot_2014}.

We consider the following \emph{continuous-action} variant of Goofspiel. 
Instead of receiving a deck of discrete bids consisting of all cards of one suit, each player has a \emph{continuous} budget that they can spend to bid on the prize card in each round.
In each round, their continuous bid is subtracted from their budget.
This can, if desired, be thought of as modeling a multi-round, multi-item, auction-like scenario with a  budget constraint for each bidder. 
To allow the players to randomize over their 1-dimensional actions (the bids), we inject their strategy networks with 1-dimensional latent input noise in addition to the observation, as described in \citet{ijcai2023p317}.

\begin{figure}[!ht]
    \centering
    \includegraphics{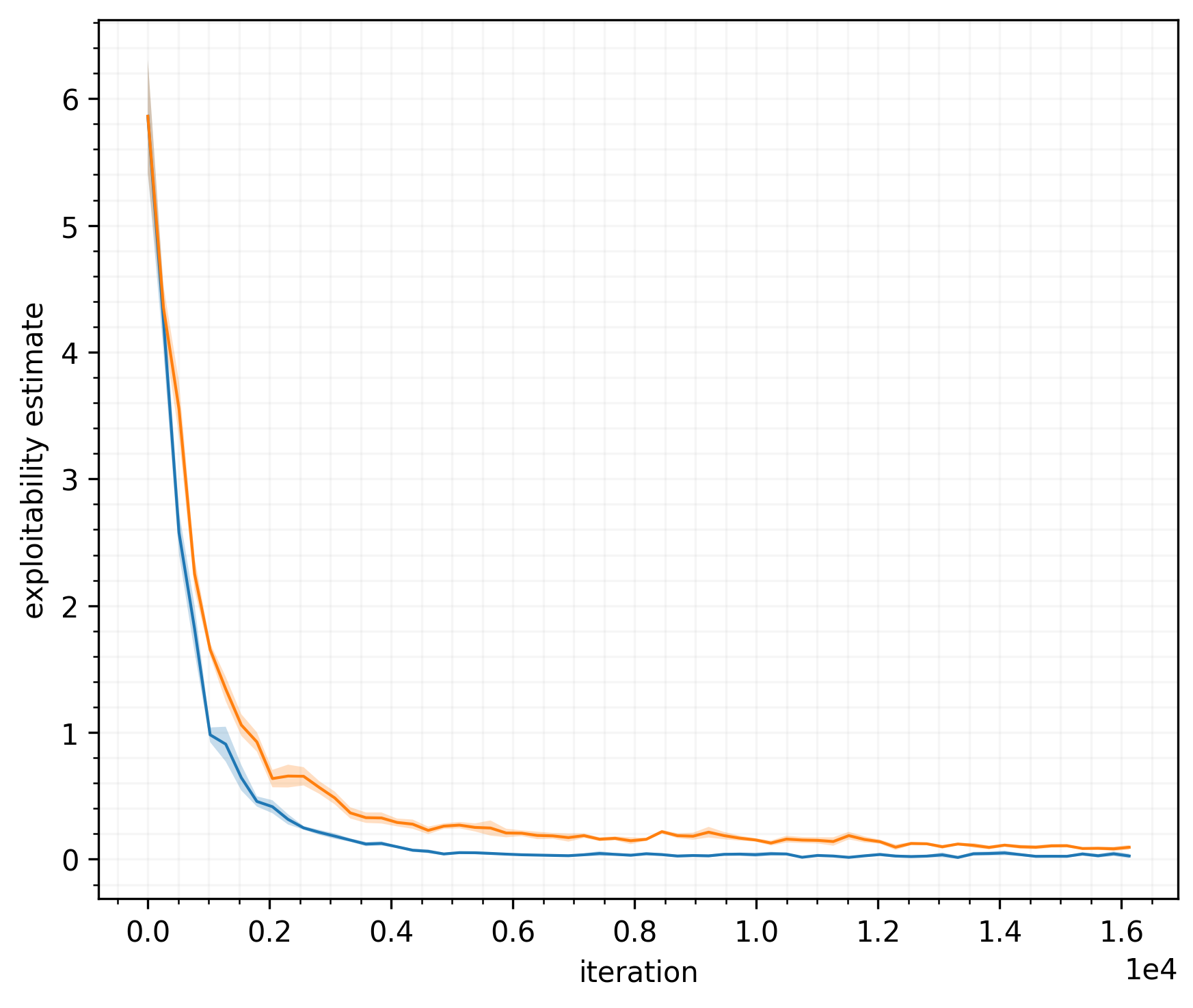}
    \includegraphics{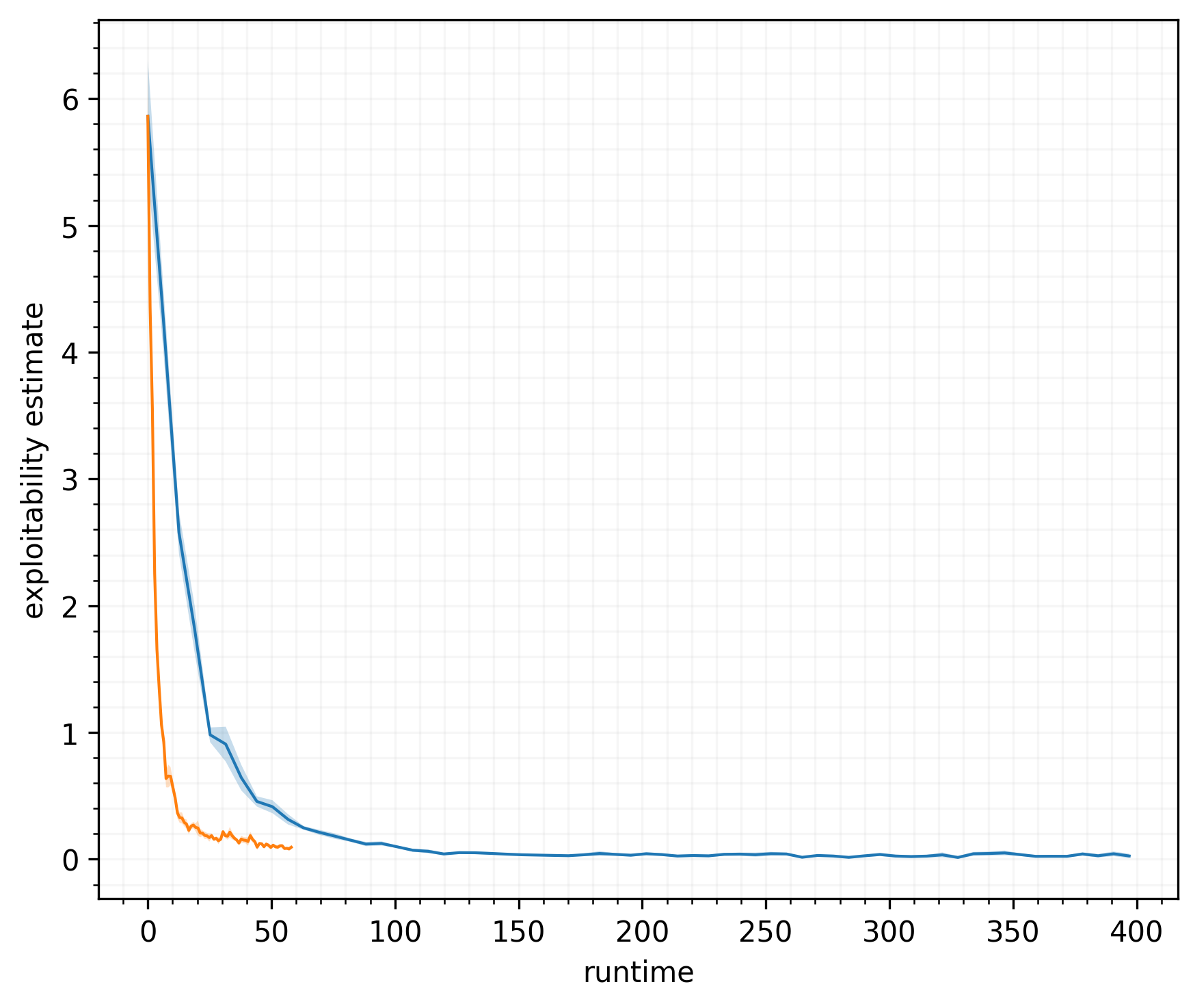}
    \caption{20-player Goofspiel.}
    \label{fig:goofspiel}
\end{figure}

Figure \ref{fig:goofspiel} shows the exploitability over the course of training on 20-player continuous-action Goofspiel.
Our method attains low exploitability with significantly faster run time than the classical method.

%% file: conclusion.tex
\section{Conclusions and future research}
\label{sec:conclusion}

We tackled the problem of computing an approximate Nash equilibrium of a game with a black-box utility function, for which we lack access to gradients.
To do this, we combined equilibrium-seeking gradient dynamics with a new zeroth-order approach to computing the simultaneous gradient.
This approach reduces the number of utility function evaluations per iteration from \emph{linear} in the number of players to \emph{constant} in the number of players.
Our method performs a \emph{joint} perturbation on all players' strategies at once, rather than perturbing each one individually.
When utility function evaluation is expensive (\emph{e.g.}, in terms of wall time, memory, or other resources), this can significantly reduce the cost of training.
We compared our approach to the standard one on several games, showing a significant reduction in training time.
For future work, we would like to test and analyze the combination of our joint-perturbation approach with alternative equilibrium-seeking dynamics, such as extragradient and optimistic gradient dynamics.

%% file: acknowledgements.tex
\section*{Acknowledgements}

This material is based on work supported by the Vannevar Bush Faculty Fellowship ONR N00014-23-1-2876, National Science Foundation grants RI-2312342 and RI-1901403, ARO award W911NF2210266, and NIH award A240108S001.

%% file: code.tex
\section{Code}

For our experiments, we used Python 3.12.3 with the following libraries:

\begin{itemize}
    \item \texttt{jax} 0.4.30 \citep{jax2018github}: \url{https://github.com/google/jax}
    \item \texttt{flax} 0.8.5 \citep{flax2020github}: \url{https://github.com/google/flax}
    \item \texttt{optax} 0.2.3 \citep{deepmind2020jax}: \url{https://github.com/google-deepmind/optax}
    \item \texttt{matplotlib} 3.9.1 \citep{Hunter:2007}: \url{https://github.com/matplotlib/matplotlib}
    \item \texttt{scipy} 1.14.0 \citep{2020SciPy-NMeth}:\url{https://github.com/scipy/scipy}
\end{itemize}

An implementation of the classical method, as well as our method, is shown below.
\\
\begin{lstlisting}
from functools import partial

import jax
import optax
from jax import numpy as jnp, random

def replace(seq, index, value):
    return seq[:index] + type(seq)([value]) + seq[index + 1 :]

def random_split_tree(key, tree):
    treedef = jax.tree.structure(tree)
    keys = random.split(key, treedef.num_leaves)
    return jax.tree.unflatten(treedef, keys)

def pgrad(f, scale, batch_size=2, dist="normal"):
    assert batch_size % 2 == 0, "batch_size must be even"
    pairs = batch_size // 2

    def pgrad_f(x, key):
        key, subkey = random.split(key)
        keys = random_split_tree(subkey, x)

        def g(key, array):
            match dist:
                case "normal":
                    z = random.normal(key, (pairs,) + array.shape)
                case "spsa":
                    z = random.rademacher(key, (pairs,) + array.shape)
                case _:
                    raise NotImplementedError(dist)
            z = jnp.concatenate([z, -z])
            return z

        zs = jax.tree.map(g, keys, x)
        xs = jax.tree.map(lambda x, zs: x + zs * scale, x, zs)

        keys = random.split(key, pairs)
        keys = jnp.concatenate([keys, keys])

        ys = jax.vmap(f)(xs, keys)

        s = 1 / (batch_size * scale)

        def h(zs):
            return jnp.tensordot(ys, zs, (0, 0)) * s

        return jax.tree.map(h, zs)

    return pgrad_f

def sim_pseudo_gradients(u, x, key, scale):
    def f(xi, key, i):
        return u(replace(x, i, xi), key)[i]

    pgrads = type(x)(
        [pgrad(partial(f, i=i), scale)(xi, key) for i, xi in enumerate(x)]
    )
    return pgrads

def joint_perturb_sim_pseudo_gradients(u, x, key, scale):
    pjac = pgrad(u, scale)(x, key)
    pjac_diag = [jax.tree.map(lambda s: s[i], si) for i, si in enumerate(pjac)]
    return pjac_diag

class StochasticGradientAscent:
    def __init__(self, fn, optimizer, grad=jax.grad):
        self.fn = fn
        self.optimizer = optimizer
        self.grad = grad

    def init(self, params, key):
        opt_state = self.optimizer.init(params)
        return params, opt_state

    def grads(self, params, key):
        return self.grad(self.fn)(params, key)

    def step(self, state, key):
        params, opt_state = state
        grads = self.grads(params, key)
        grads = jax.tree.map(jnp.negative, grads)
        updates, opt_state = self.optimizer.update(grads, opt_state, params)
        params = optax.apply_updates(params, updates)
        params = self.project_params(params)
        return (params, opt_state), state

    def get_params(self, state):
        params, opt_state = state
        return params

    def project_params(self, params):
        return params

class SimPseudoGrad(StochasticGradientAscent):
    """Simultaneous pseudo-gradient ascent."""

    def __init__(self, fn, optimizer, scale, joint_perturb=False):
        self.fn = fn
        self.optimizer = optimizer
        self.scale = scale
        self.joint_perturb = joint_perturb

    def grads(self, x, key):
        if self.joint_perturb:
            return joint_perturb_sim_pseudo_gradients(
                self.fn, x, key, self.scale
            )
        else:
            return sim_pseudo_gradients(self.fn, x, key, self.scale)
\end{lstlisting}